\documentclass[a4paper,11pt]{article}
\pdfoutput=1 
\usepackage{booktabs}
\usepackage{soul}
\usepackage{graphicx}
\usepackage{caption}
\usepackage[titletoc]{appendix} 
\usepackage{subcaption}
\usepackage{bm}
\usepackage{jheppub} 
\usepackage[T1]{fontenc} 
\title{\boldmath The bottomonium spectrum at finite temperature from $N_f=2+1$ lattice QCD}
\author[a]{G. Aarts,}
\author[a]{C. Allton,}
\author[b]{T. Harris,}
\author[c]{S. Kim,}
\author[d]{M. P. Lombardo,}
\author[b]{S. M. Ryan,}
\author[e]{ and J.-I. Skullerud}
\affiliation[a]{Department of Physics, College of Science, Swansea University, Swansea, United Kingdom}
\affiliation[b]{School of Mathematics, Trinity College, Dublin 2, Ireland}
\affiliation[c]{Department of Physics, Sejong University, Seoul 143-747, Korea}
\affiliation[d]{INFN-Laboratori Nazionali di Frascati, I-00044, Frascati (RM) Italy}
\affiliation[e]{Department of Mathematical Physics, National University of Ireland Maynooth, Maynooth, County Kildare, Ireland}
\emailAdd{g.aarts@swan.ac.uk}
\emailAdd{c.allton@swansea.ac.uk}
\emailAdd{tharris@tcd.ie}
\emailAdd{skim@sejong.ac.kr}
\emailAdd{Mariapaola.Lombardo@lnf.infn.it}
\emailAdd{ryan@maths.tcd.ie}
\emailAdd{jonivar@thphys.nuim.ie}
\abstract{We present results on the bottomonium spectrum at temperatures above and below the deconfinement crossover temperature, $T_c$, from dynamical lattice QCD simulations.
The heavy quark is treated with a non-relativistic effective field theory on the lattice and serves as a probe of the hot medium.
Ensembles with a finer spatial lattice spacing and a greater range of temperatures below $T_c$ than those previously employed by this collaboration are used.
In addition, there are $N_f=2+1$ flavours of Wilson clover quark in the sea with $M_\pi\approx400$ MeV and we perform a more careful tuning of the bottom quark mass in this work.
We calculate the spectral functions of S and P~wave bottomonium states using the maximum entropy method and confirm earlier findings on the survival of the ground state S~wave states up to at least $2T_c$ and the immediate dissociation of the P~wave states above $T_c$.
}
\begin{document} 
\maketitle
\flushbottom
\section{Introduction}
\label{sec:intro}
The dissociation of heavy quarkonia in a deconfined medium may provide a valuable thermometer for relativistic heavy-ion collisions~\cite{Brambilla:2010cs}.
Dissociation, or melting, contributes to the suppression of the yield of these states in nuclear collisions compared with hadronic ones~\cite{Matsui:1986dk}.
Suppression patterns~\cite{Karsch:1987pv,Karsch:2005nk} are complicated by the statistical recombination of heavy quarks in the quark-gluon plasma (QGP).
However, competing effects are expected to be less pronounced for bottomonium than for charmonium~\cite{Rapp:2008tf}.
Indeed sequential suppression has been observed recently in the $\Upsilon$ system by CMS at the LHC~\cite{Chatrchyan:2012lxa}.
It is therefore desirable to understand the dissociation of mesons in the bottomonium system from ab initio QCD.
Analytic results from effective field theories~\cite{Laine:2006ns,Laine:2007gj,Burnier:2007qm,Beraudo:2007ky,Brambilla:2010vq,Brambilla:2011sg} can aid the identification and interpretation of the relevant physical phenomena, such as the role of the finite width of such states in the plasma as well as the familiar colour-Debye screening mechanism~\cite{Burnier:2007qm}.
However, in order to achieve a separation of scales, they generally rely on a choice of the parametric size of the temperature, for instance $T\sim gm_b$ or $T\sim g^2m_b$, where $m_b$ is the heavy quark mass and $g$ is the coupling which is assumed to be sufficiently weak for a hierarchy of scales to emerge. 
In contrast, by directly simulating the non-relativistic field theory for the heavy quark (NRQCD) non-perturbatively, all that is required is that $m_b\gg T$, which is certainly satisfied for the bottom quark in the relevant regime up to $4$ or $5T_c$.
Numerical lattice simulations are well suited to investigate the properties of the strongly-coupled QGP and to capture the non-perturbative dynamics of the hot medium formed around the crossover temperature, $T_c$.
However, it remains a challenge to control systematic errors in dynamical lattice simulations.
This paper continues the work of the \textsc{fastsum} collaboration's programme to investigate phenomenologically relevant observables in the QGP with improved control over uncertainties~\cite{Aarts:2010ek,Aarts:2011sm,Aarts:2012ka,Aarts:2013kaa,Amato:2013naa}.
Here, we calculate bottomonium spectral functions from $N_f=2+1$ ensembles using the maximum entropy method. 
The determination of the spectral functions from the lattice aims to complement other approaches such as weak-coupling effective field theory and potential models to give a complete understanding of the behaviour of heavy quarkonium through the deconfinement crossover.
Previous studies of quarkonium spectral functions from lattice QCD, mostly  for charmonium, include \cite{Asakawa:2003re,Datta:2003ww,Jakovac:2006sf,Aarts:2007pk,Ding:2012sp,Borsanyi:2014vka}. 
An earlier analysis of the spectral functions by this collaboration from $N_f=2$ flavour ensembles suggested the ground state S~waves ($\eta_b$ and $\Upsilon$ channels) survive up to at least $2T_c$ while the first excited state may be suppressed in the deconfined phase, both at zero~\cite{Aarts:2011sm} and non-zero momentum~\cite{Aarts:2012ka}.
Those conclusions are consistent with predictions from effective field theory~\cite{Burnier:2007qm} and experimental data~\cite{Chatrchyan:2012lxa}, although a detailed comparison with the latter is beyond the scope of this paper.
An examination of the correlation functions indicated~\cite{Aarts:2010ek} that the P~waves ($h_b, \chi_{b1,2,3}$ channels) are greatly modified directly above $T_c$.
By comparison with the expectations from the free continuum effective theory the observed thermal modification of the correlators provided evidence in favour of the hypothesis that the P~waves dissociate above the crossover temperature.
This interpretation is supported by the examination of the spectral functions in those channels~\cite{Aarts:2013kaa}.
Here, we extend that work by using new $N_f=2+1$ ensembles with improved lattice parameters.
The results for the spectral functions are compatible with earlier findings but in the analysis of the correlators we note that the finite threshold plays a more prominent role, discussed in section~\ref{sec:corrs}.
The following section outlines the simulation details and zero temperature calibration. We present the bottomonium correlators and spectral functions at finite temperature in section~\ref{sec:finiteTnrqcd}, discuss systematic effects in the spectral function reconstruction in section~\ref{sec:systests} and conclude in section~\ref{sec:concs}.
\section{Lattice set-up}
\label{sec:lattice}
In this work we employ ensembles with anisotropic lattice spacings, using a tadpole-improved Wilson clover quark action for the light and strange quarks and a tadpole- and Symanzik-improved gauge action.
Tree-level improvement coefficients are used for both the fermion and gauge actions.
The tuning of the lattice parameters was performed by the Hadron Spectrum Collaboration and further discussion can be found in ref.~\cite{Edwards:2008ja}.
A range of temperatures above and below the deconfinement crossover is accessible from $0.76T_c$ to $1.90T_c$.
The fixed-scale approach is used whereby the temperature is varied by changing the number of temporal lattice sites while the lattice spacings are kept constant.
This reduces the overhead of zero temperature simulations required to tune the lattice parameters.
The renormalized anisotropy, $\xi\equiv a_s/a_\tau$, is tuned to $\xi=3.5$, which allows us to maintain an adequate resolution of the correlation functions to obtain the spectral functions at high temperatures.
\begin{table}
    \centering
    \begin{tabular}{r*{8}{c}}
        \toprule
        $N_s$            & 16  & 24   & 24   & 24   & 24   & 24   & 24   & 24   \\
        $N_\tau$         & 128 & 40   & 36   & 32   & 28   & 24   & 20   & 16   \\
        \midrule
        $T/T_c$          & $\sim0$  & 0.76 & 0.84 & 0.95 & 1.09 & 1.27 & 1.52 & 1.90 \\
        $T$ (MeV)        & $\sim0$  & 141  & 156  & 176  & 201  & 235  & 281  & 352  \\
        $N_\textrm{cfg}$ & 499 & 502  & 503  & 998  & 1001 & 1002 & 1000 & 1042 \\
        \bottomrule
    \end{tabular}
    \caption{Summary of the ensembles used in this work.
        The crossover temperature is determined from the renormalized Polyakov loop~\cite{Allton:2014uia}.
        The zero temperature tuning of the lattice parameters  was completed by the Hadron Spectrum Collaboration~\cite{Edwards:2008ja}.}
    \label{tab:ensem_temp}
\end{table}
These ``second generation'' ensembles~\cite{Amato:2013naa} represent multiple improvements over the ``first generation'' ones used in the previous work by this collaboration~\cite{Aarts:2010ek,Aarts:2011sm,Aarts:2012ka,Aarts:2013kaa}.
In particular, the spatial lattice spacing, $a_s=0.1227(8)\,$fm is finer, while $M_\pi/M_\rho\sim0.45$ is reduced.
Moreover, the strange quark is now included in the sea.
Details are given in tables~\ref{tab:ensem_temp} and~\ref{tab:latparam}.
On these ensembles the pion is relatively heavy, $M_\pi\approx 400$~MeV, while the kaon is roughly physical, $M_K\approx500$~MeV~\cite{Lin:2008pr}.
The highest accessible temperature is slightly reduced in this study with respect to the earlier work.
\begin{table}
    \centering
    \begin{tabular}{cccccccc}
        \toprule
                           & $N_f$ & $a_s$ (fm)        & $a_\tau^{-1}$ (GeV)& $\xi$ & $M_\pi/M_\rho$ & $a_sm_b$&$E_0$ (MeV) \\
         \midrule
         First generation  & 2     & 0.162     & 7.35      & 6     & 0.54           & 4.5    & 8570  \\
         Second generation & $2+1$ & 0.1227(8) & 5.63(4)      & 3.5   & 0.45           & 2.92   & 8252(9) \\
        \bottomrule
    \end{tabular}
    \caption{Comparison between lattice parameters used in earlier work~\cite{Morrin:2006tf,Oktay:2010tf,Aarts:2010ek,Aarts:2011sm,Aarts:2012ka,Aarts:2013kaa} (first generation) and this work (second generation).
            $E_0$ is the additive shift in the bottomonium energy, see eq.~(\ref{eq:shift}).}
    \label{tab:latparam}
\end{table}
\subsection{Lattice NRQCD}
\label{ssec:latticenrqcd}
NRQCD is an effective field theory with power counting in the heavy quark velocity in the bottomonium rest frame, $v\sim |\bm p|/m_b$~\cite{Lepage:1992tx}.
The finite lattice spacing cuts off the relativistic modes of the heavy quark in the discretized theory.
The heavy quark and anti-quark fields decouple and their numbers are separately conserved.
Their propagators, $S(x)$, solve an initial-value problem whose discretization leads to the following choice for the evolution equation 
\begin{align}
    S(x+a_\tau\bm e_\tau)=
        \left(1-\frac{a_\tau H_0|_{\tau+a_\tau}}{2k}\right)^kU_\tau^\dagger(x)
        \left(1-\frac{a_\tau H_0|_{\tau}}{2k}\right)^k\left(1-{a_\tau\delta H}\right)
        S(x),
\end{align}
with $U_\tau(x)$ being the temporal gauge link at site $x$ and $\bm e_\tau$ the temporal unit vector. The leading-order Hamiltonian is defined by
\begin{align}
    H_0=-\frac{\Delta^{(2)}}{2m_b},
        \qquad\textrm{with}\qquad\Delta^{(2n)}
        =\sum_{i=1}^3\left(\nabla_i^+\nabla_i^-\right)^{n}.
\end{align}
The higher order covariant finite differences are written in terms of the components of the usual forward ($\nabla^+_i$) and backward ($\nabla^-_i$) first order ones.
The correction terms are given by
\begin{align}
    \label{eq:deltaH}
    \begin{split}
        \delta H=&-\frac{\left(\Delta^{(2)}\right)^2}{8m_b^3}
        +\frac{ig_0}{8m_b^2}\left(\bm\nabla^\pm\cdot \bm E
        - \bm E\cdot\bm\nabla^\pm\right)\\
        &-\frac{g_0}{8m_b^2}\bm\sigma\cdot\left(\bm\nabla^\pm\times \bm E
        - \bm E\times\bm\nabla^\pm\right)-\frac{g_0}{2m_b}\bm \sigma\cdot\bm B\\
        &+\frac{a_s^2\Delta^{(4)}}{24m_b}
        -\frac{a_\tau\left(\Delta^{(2)}\right)^2}{16k m_b^2},
    \end{split}
\end{align}
which incorporate $O(v^4)$ corrections as well as the leading spin-dependent corrections.
The matching coefficients are determined at tree-level.
The terms on the final line of eq.~(\ref{eq:deltaH}) remove the $O(a_s^2)$ errors in $H_0$ and the $O(a_\tau)$ errors of the evolution equation respectively.
The choice of $k=1$ for Lepage's parameter is sufficient for these anisotropic lattices in order to satisfy the stability criterion $\max|1-a_\tau H_0/2k|<1$.
Other choices of $k=2,3$ were investigated but their effects are negligible.
The clover definition of the field-strength tensor is used to define the unimproved chromoelectric and magnetic fields and $\bm \nabla^\pm$ is the symmetric covariant finite difference operator.
Tadpole improvement is implicit and implemented by dividing all links by the mean link determined from the fourth root of the average plaquette.
\begin{figure}[t]
    \centering
    \includegraphics[scale=0.9]{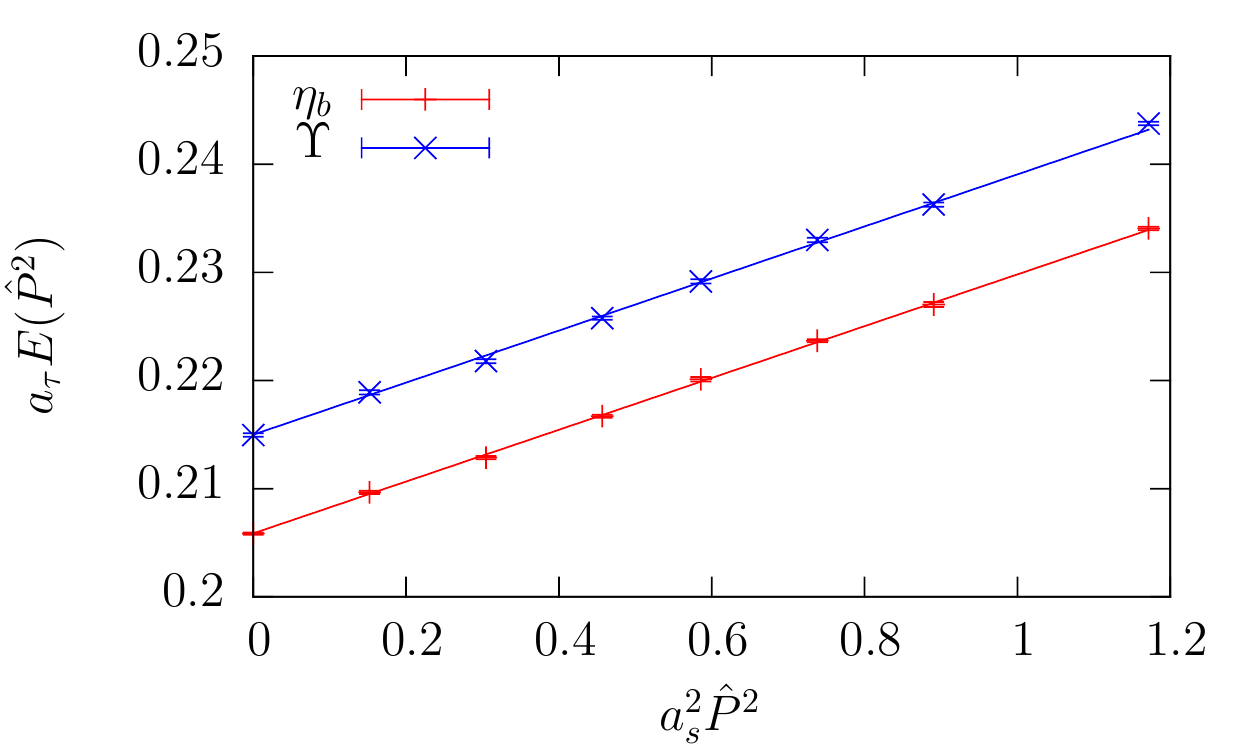}
    \caption{Zero temperature dispersion relations in the $\eta_b$ and $\Upsilon$ channels used to determine the 1S spin-averaged kinetic mass, $M_2(\overline{\textrm{1S}})$, with fits given in eq.~(\ref{eq:dispfit}).}
    \label{fig:dispersion}
    \label{fig:corrsdisp}
\end{figure}
Only energy differences are physically significant in NRQCD because the rest-mass energy can be removed from the heavy quark dispersion relation by performing a field transformation.
The determination of the lattice heavy quark mass, $a_s m_b$, is achieved by tuning a hadronic kinetic mass, $M_2$, at zero temperature which appears in the hadronic lattice dispersion relation
\begin{align}
    \label{eq:hadron_disp}
    \begin{split}
    a_\tau E(\hat P^2)&=a_\tau E(0)+\frac{a_s^2\hat P^2}{2\xi^2a_\tau M_2}+\ldots\\
        a_s^2 \hat P^2&=4\sum_{i=1}^3\sin^2\left(\frac{\pi n_i}{N_s}\right).
    \end{split}
\end{align}%
where $n_i={-N_s/}{2}+1,\ldots,{N_s/}{2}$. In this work we tune the heavy quark mass by requiring the spin-averaged $1$S kinetic mass, $M_2(\overline{1\textrm{S}})=(M_2(\eta_b)+3M_2(\Upsilon))/4$, to be equal to its experimental value.
Using the spin-averaged kinetic mass mitigates the systematic error due to the determination of the hyperfine splitting in the kinetic mass~\cite{Dowdall:2011wh}.
The fits to the dispersion relations, shown in figure~\ref{fig:dispersion}, at the tuned value of the heavy quark mass $a_sm_b=2.92$, are given by
\begin{align}
    \begin{split}
        a_\tau E(\eta_b)&=0.2058(2)+0.0239(3)a_s^2\hat P^2\\
        a_\tau E(\Upsilon)&=0.2150(3)+0.0241(3)a_s^2\hat P^2\\
    \end{split}
    \label{eq:dispfit}
\end{align}
with a statistical error determined from a bootstrap analysis.
Higher order terms in the dispersion relation could not be resolved within the statistical precision.
The tuned value of the heavy quark mass corresponds to $M_2(\overline{1\textrm S})=9560(110) \textrm{ MeV}$ which is consistent with the experimental value, $M_\mathrm{expt}(\overline{\mathrm{1S}})=9444.7(8)$ MeV. %
The error includes a contribution from the statistical uncertainty in the scale set from the $\Omega$ baryon mass~\cite{Moir:2013ub}.
This tuning is an improvement over the ad hoc choice of the heavy quark mass in the previous study.
\subsection{Zero temperature results}
\label{sec:zeroT}
\begin{figure}[t]
    \centering
    \includegraphics[scale=0.9]{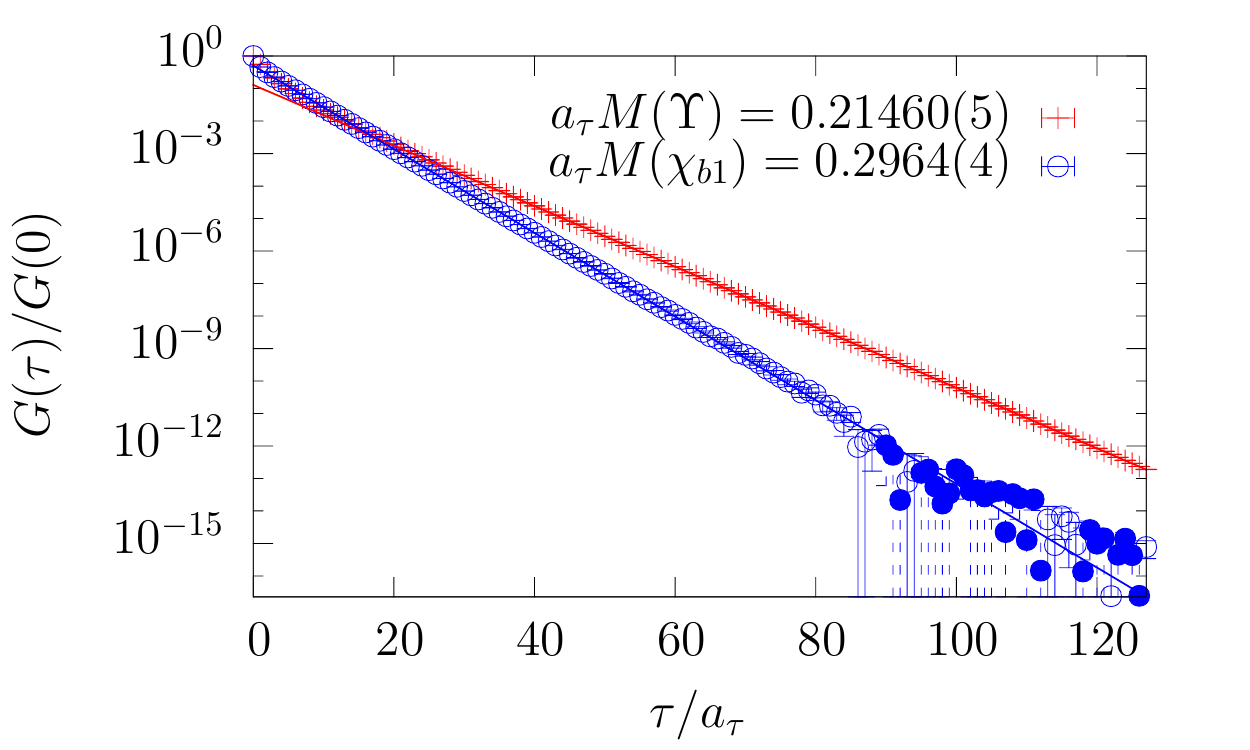}
    \caption{Correlation functions in the $\Upsilon$ and $\chi_{b1}$ channels with the corresponding ground state energies extracted from exponential fits.
        Filled symbols denote negative data excluded from the fit.}
            \label{fig:corrs}
\end{figure}

\begin{figure}[t]
    \centering{
        \begin{subfigure}[t]{0.5\textwidth}
            \includegraphics[width=\textwidth]{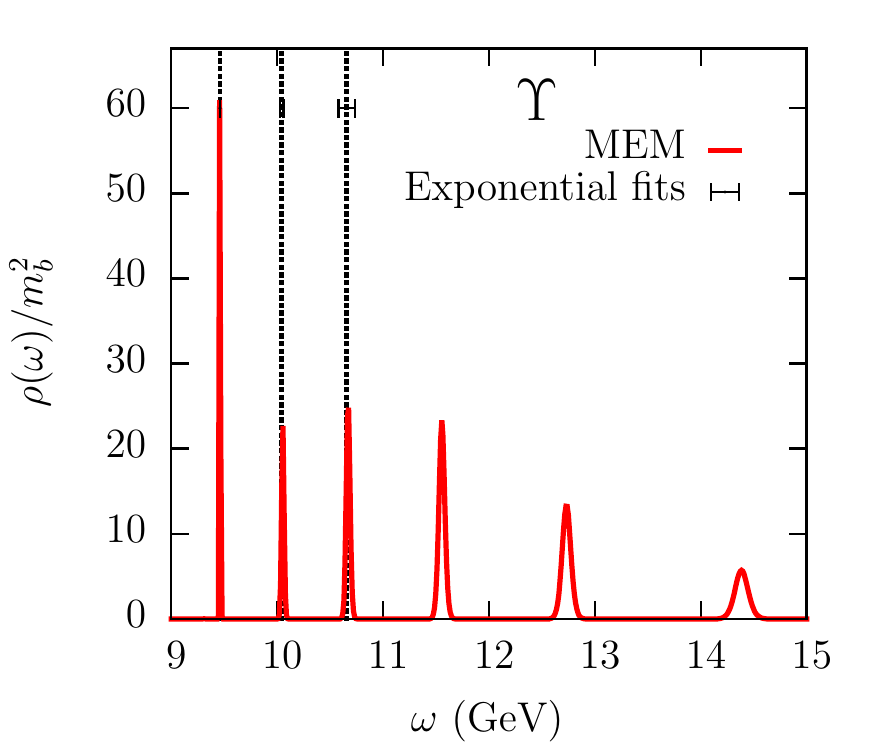}
        \end{subfigure}%
        \begin{subfigure}[t]{0.5\textwidth}
            \includegraphics[width=\textwidth]{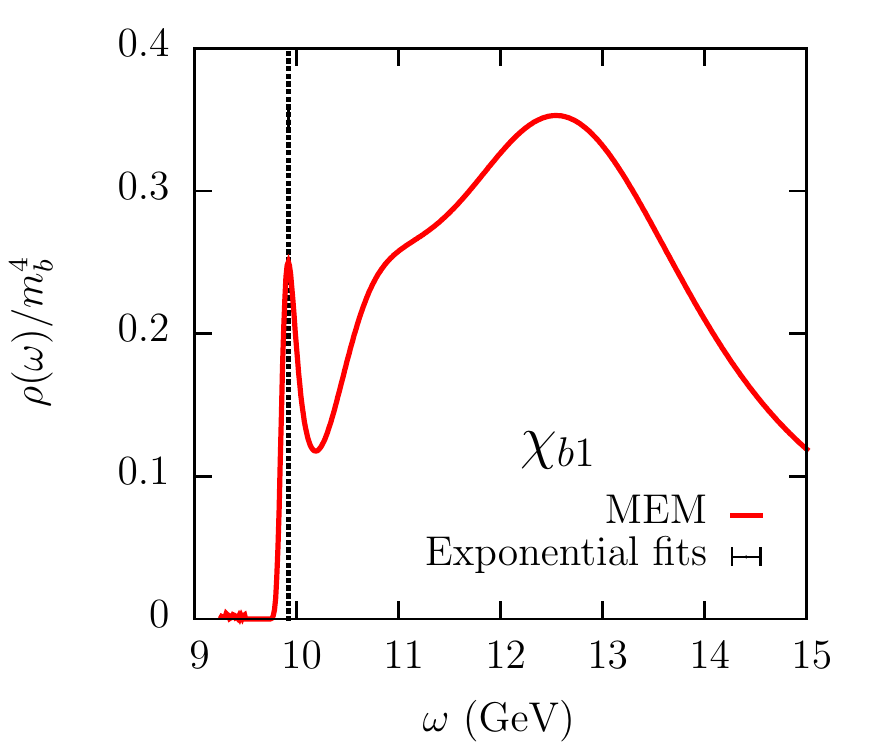}
        \end{subfigure}
    }
    \vskip -1em
    \caption{Spectral functions in the $\Upsilon$ (left) and $\chi_{b1}$ (right) channels at zero temperature with energies determined from exponential fits shown as black dotted lines with statistical errors.}
    \label{fig:rho_16x128}
\end{figure}

Figure~\ref{fig:corrs} shows correlation functions in the $\Upsilon$ (red crosses) and $\chi_{b1}$ (blue circles) channels and the energies determined from single exponential fits.
The experimental $\Upsilon(1\mathrm S)$ mass is used to fix the absolute energy shift, $E_0$, of spectral quantities 
\begin{align}
    E_0=M_\mathrm{expt}(\Upsilon)-M(\Upsilon)=8252(9)\,\textrm{MeV}.
    \label{eq:shift}
\end{align}
The resulting zero temperature spectrum is given in table~\ref{tab:spectrum}.
We note that in ref.~\cite{Detmold:2012pi} a heavy quark action with O($v^6$) corrections was used in conjunction with the same zero temperature ensemble used in this work.
Slight discrepancies between the spectra could arise from a different heavy quark action including a different choice for the matching coefficients.
Precision studies of the zero temperature bottomonium spectrum using NRQCD have been performed by the HPQCD collaboration~\cite{Dowdall:2011wh,Dowdall:2013jqa}.
\begin{table}[ht]
    \centering
    \begin{tabular}{c*{4}{c}}
        \toprule
        $n^{S+1}L_J$              & State       & $a_\tau M$ & $E_0+M$ (MeV) & $M_\mathrm{expt}$ (MeV) \\ 
        \midrule
        $1^1\textrm S_0$          & $\eta_b$    & 0.20549(4)                & 9409(12)                    & 9398.0(3.2)             \\ 
        $2^1\textrm S_0$          & $\eta_b'$   & 0.311(3)                  & 10004(21)                  & 9999(4)                 \\ 
        $1^3\textrm S_1$          & $\Upsilon$  & 0.21460(5)                & 9460*                      & 9460.30(26)             \\ 
        $2^3\textrm S_1$          & $\Upsilon'$ & 0.318(3)                  & 10043(22)                  & 10023.26(31)            \\ 
        $1^1\textrm P_1$          & $h_b$       & 0.2963(4)                 & 9920(15)                   & 9899.3(1.0)             \\ 
        $1^3\textrm P_0$          & $\chi_{b0}$ & 0.2921(4)                 & 9896(15)                   & 9859.44(52)             \\ 
        $1^3\textrm P_1$          & $\chi_{b1}$ & 0.2964(4)                 & 9921(15)                   & 9892.78(40)             \\ 
        $1^3\textrm P_2$          & $\chi_{b2}$ & 0.2978(4)                 & 9928(15)                   & 9912.21(40)             \\ 
        \bottomrule
    \end{tabular}
    \caption{Bottomonium spectrum from standard exponential fits where the $\Upsilon$ mass has been used to set the absolute energy shift, $E_0$.
        The error quoted in the third column is statistical, while the error in the fourth column includes a contribution from the statistical uncertainty in the scale~\cite{Lin:2008pr,Moir:2013ub}.
        The experimental masses are taken from the Particle Data Group booklet~\cite{Beringer:1900zz}.
    }
    \label{tab:spectrum}
\end{table}
The maximum entropy method (MEM) with Bryan's algorithm~\cite{Asakawa:2000tr} was used to obtain the spectral functions, $\rho(\omega)$, from the hadronic correlation functions, which are related through
\begin{align}
    G(\tau)=\int_{\omega_\mathrm{min}}^{\omega_\mathrm{max}} \frac{\mathrm d\omega}{2\pi}\,K(\tau,\omega)\,\rho(\omega),
    \qquad K(\tau,\omega)=e^{-\omega\tau}.
    \label{eq:specfn}
\end{align}
Although this is a typical ill-posed problem given the discrete and noisy estimator for the correlation function, $G(\tau)$, MEM gives a unique solution after specification of the default model.
Further details on the implementation used can be found in ref.~\cite{Aarts:2011sm}, where also the default model dependence is studied in detail.
Figure~\ref{fig:rho_16x128} shows the spectral functions in the S~wave $\Upsilon$ (left) and P~wave $\chi_{b1}$ (right) channels along with energies extracted from multi-exponential fits directly to the correlators using the \textsc{CorrFitter} package~\cite{Hornbostel:2011hu} shown with vertical black dotted lines.
Good agreement is observed between the energies extracted directly from the correlators and the peak positions in the reconstructed spectral functions.
We note that while as many as six well-defined peaks can be discerned in the S~wave channel presented here, the third and higher peaks are not compatible with the experimental spectrum~\cite{Beringer:1900zz}.
A more thorough investigation of the zero temperature spectrum could be undertaken following the HPQCD approach~\cite{Dowdall:2011wh,Dowdall:2013jqa}.
\section{Bottomonium at finite temperature}
\label{sec:finiteTnrqcd}
The heavy quarks are valence quarks which propagate through the thermal medium according to the solution of the initial value problem in NRQCD.
Their propagators do not satisfy anti-periodic thermal boundary conditions so the heavy quarks are not in thermal equilibrium with the medium.
This is illustrated in the representation of the correlation function in eq.~(\ref{eq:specfn}) which is manifestly not symmetric about $\tau=1/2T$, while the kernel, $K(\tau,\omega)$, is independent of the temperature.
The thermal modification of the correlators can therefore be directly attributed to the modification of the associated spectral function.
The asymmetry of the hadronic correlation functions can be seen explicitly in figure~\ref{fig:corrs}.
These simplifications result from replacing $\omega\rightarrow2m_b+\omega$ and taking the $m_b/T\rightarrow\infty$ limit in the standard kernel at finite temperature
\begin{align}
    K(\tau,\omega) = \frac{\cosh(\omega\tau-\omega/2T)}{\sinh(\omega/2T)} \rightarrow
    e^{-\omega\tau}(1+n_B(\omega)) +  e^{\omega\tau} n_B(\omega),
\end{align}
where $n_B(\omega)= 1/(\exp(\omega/T)-1)$ is the Bose-Einstein distribution.
The inversion of eq.~(\ref{eq:specfn}) to obtain the spectral function is simpler with this asymmetric kernel, since the correlation functions are accessible to much greater temporal separations than for relativistic quarks.
Moreover, unlike a formulation in which the quarks are in equilibrium with the medium there is no constant contribution to the hadronic correlation functions.
This reflects the fact that NRQCD is an effective field theory around the two-quark threshold.
\subsection{Thermal modification of correlation functions}
\label{sec:corrs}
\begin{figure}[t]
    \centerline{
        \begin{subfigure}[t]{0.5\textwidth}
            \includegraphics[width=\textwidth]{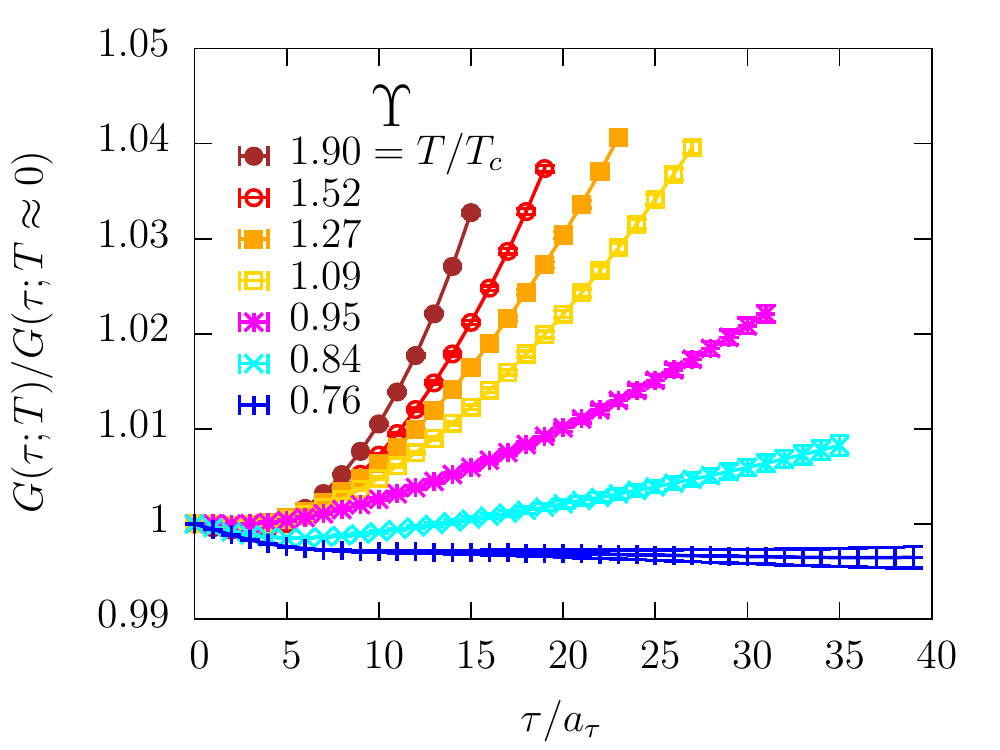}
        \end{subfigure}
        \begin{subfigure}[t]{0.5\textwidth}
            \includegraphics[width=\textwidth]{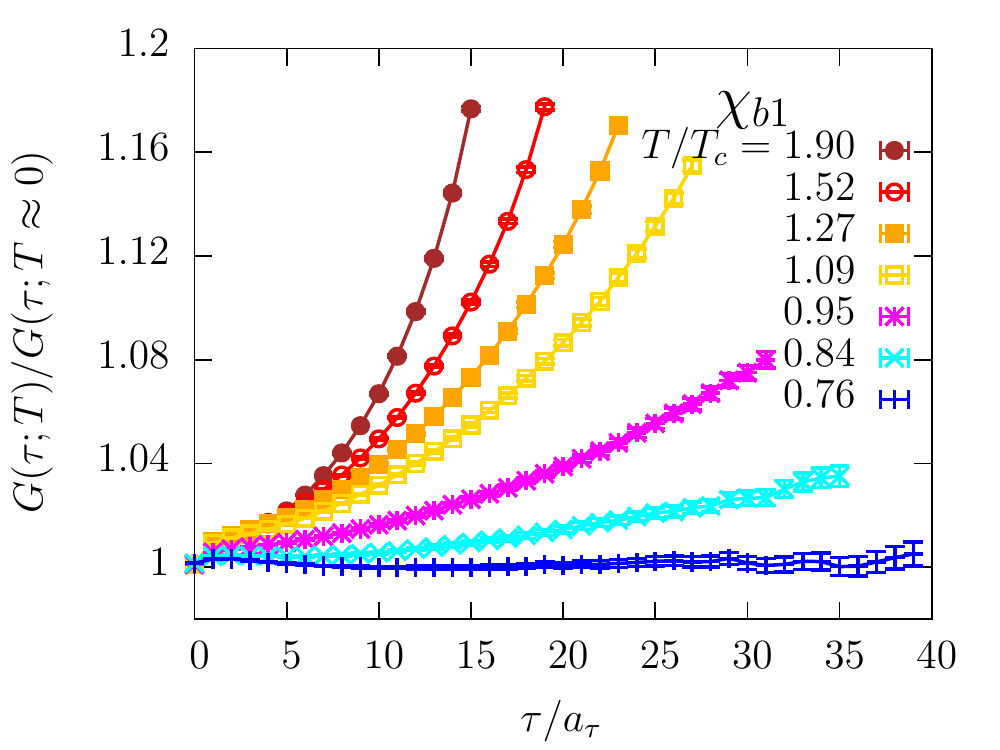}
        \end{subfigure}
    }\vskip -1em
    \caption{Thermal modification, $G(\tau;T)/G(\tau;T\approx0)$, of the correlation functions in the $\Upsilon$ (left) and $\chi_{b1}$ (right) channels.}
    \label{fig:corrs_Tdep}
\end{figure}

\begin{figure}[t]
    \centerline{
        \begin{subfigure}[t]{0.5\textwidth}
              \includegraphics[width=\textwidth]{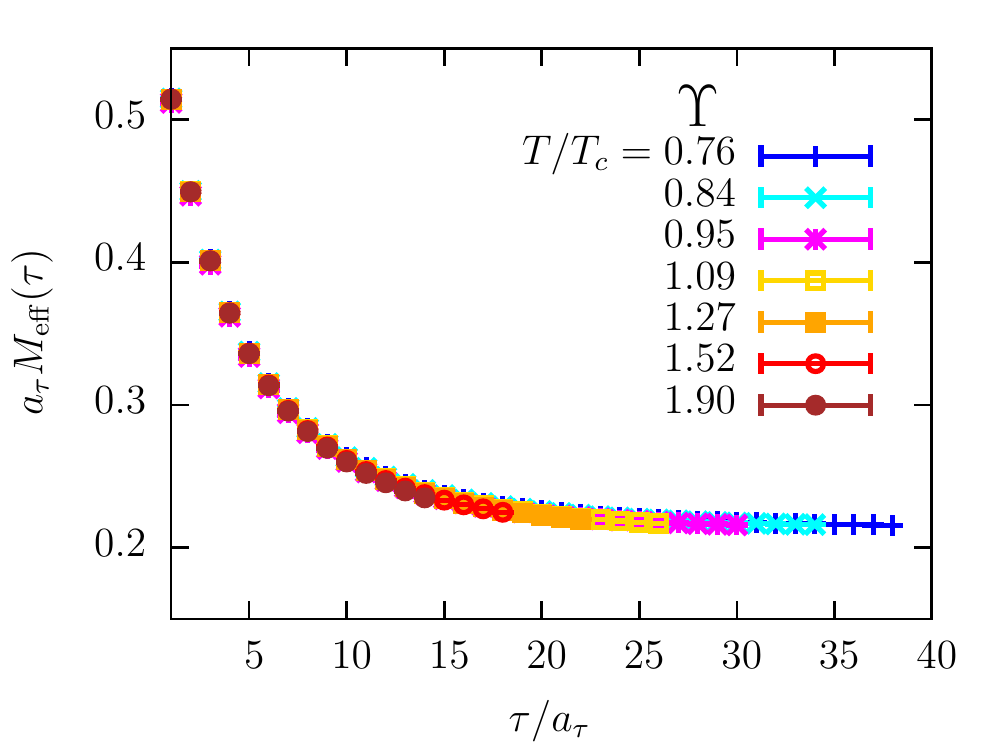}
        \end{subfigure}
        \begin{subfigure}[t]{0.5\textwidth}
              \includegraphics[width=\textwidth]{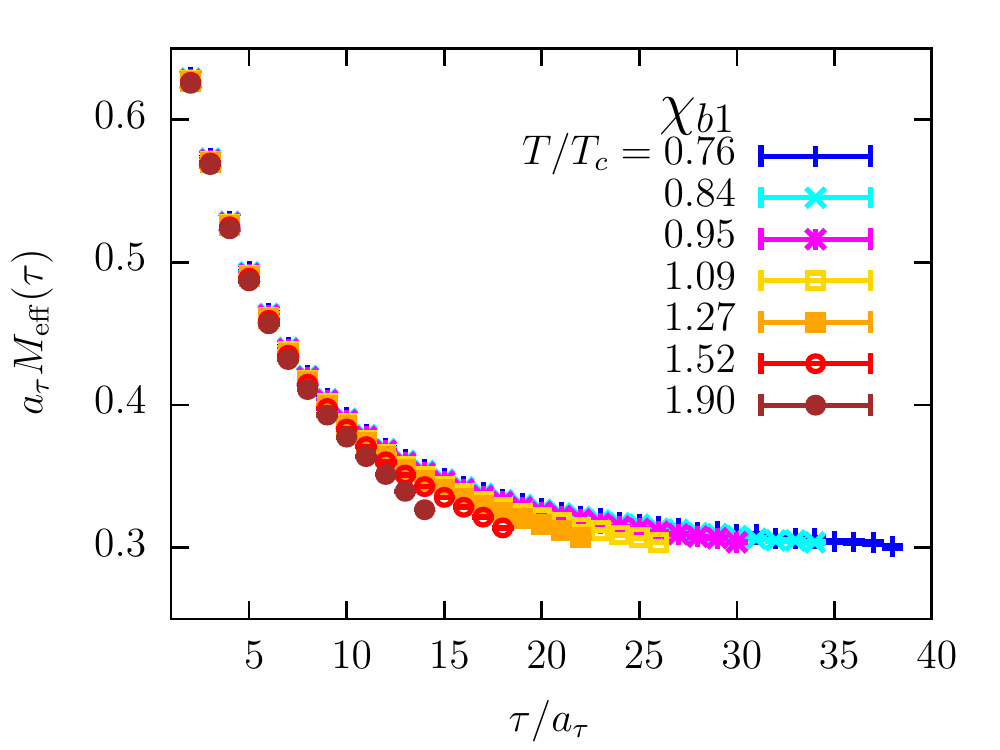}
        \end{subfigure}
    }
    \caption{Temperature dependence of the effective mass in the $\Upsilon$ (left) and the $\chi_{b1}$ (right) channels.}
    \label{fig:meff_Tdep}
\end{figure}

The ratios of the correlation functions at finite temperature to those at zero temperature are shown in figure~\ref{fig:corrs_Tdep}.
With increasing temperature the correlators are enhanced at earlier temporal separations.
We observe temperature dependence already below $T_c$ and stronger dependence above $T_c$.
An enhancement of approximately 4\% is seen in the S~wave $\Upsilon$ channel (left) above $T_c$ but there is a greater effect of almost 20\% in the P~wave $\chi_{b1}$ channel (right).
The enhancement relative to the low temperature correlators in the $N_f=2$ studies~\cite{Aarts:2011sm,Aarts:2013kaa} was approximately $2\%$ and $25\%$ in the S~wave and P~wave channels, respectively.
The correlators in the other S~wave ($\eta_b$) and P~wave ($h_b,\chi_{b0},\chi_{b2}$) channels show analogous modifications to the $\Upsilon$ and $\chi_{b1}$ channels.
The S~wave effective mass displays little temperature dependence (figure~\ref{fig:meff_Tdep}, left) but a clear effect is seen in the P~wave channel effective mass (right).
In ref.~\cite{Aarts:2010ek} it was also observed that the S~wave effective mass showed little variation with temperature while the temperature dependence in the P~wave channel effective mass was even more pronounced than visible here.
It is useful to compare the effective mass at high temperatures with that in the non-interacting infinite temperature limit.
In continuum NRQCD the spectral functions are known for free heavy quarks~\cite{Burnier:2007qm}, and are given by
\begin{align}
    \rho_{\textrm{free}}(\omega)\propto(\omega-\omega_0)^{\alpha}\,\Theta(\omega-\omega_0),
        \quad\textrm{where}\quad
        \alpha=
        \begin{cases}
            1/2, &\textrm{S~wave}.\\
            3/2, &\textrm{P~wave}.
        \end{cases}
    \label{eq:rhofree}
\end{align}
To facilitate the comparison with the interacting effective theory we have included a threshold, $\omega_0$, to account for the additive shift in the quarkonium energies%
\footnote{For free quarks the threshold occurs at $2m_b$, which within NRQCD corresponds to $\omega_0=0$.}.
This threshold may depend on the lattice parameters and the temperature.
In the infinite temperature limit the correlation functions then have the following behaviour
\begin{align}
    G_{\mathrm{free}}(\tau)\propto \frac{e^{-\omega_0\tau}}{\tau^{\alpha+1}},
    \label{eq:powerlaw}
\end{align}
and the effective mass becomes
\begin{align}
    M_\mathrm{eff}(\tau) \equiv -\frac{1}{G(\tau)}\frac{dG(\tau)}{d\tau}
    \quad\stackrel{G=G_\mathrm{free}}{\longrightarrow}\quad
    \omega_0+\frac{\alpha+1}{\tau}.
    \label{eq:effmass}
\end{align}
Pure power law decay in the correlator or the absence of plateaus in the effective mass may be less evident in the presence of a larger threshold.
In the earlier $N_f=2$ studies comparisons between the correlators and effective mass and their non-interacting counterparts were possible without including a finite threshold.
However, the threshold appears to play a more significant role here which is reflected in the fact that there is a change in the energy shift, $E_0$, of about 300 MeV between the two studies, see table~\ref{tab:latparam}.
 
\begin{figure}[t]
    \centerline{
        \begin{subfigure}[t]{0.5\textwidth}
            \includegraphics[width=\textwidth]{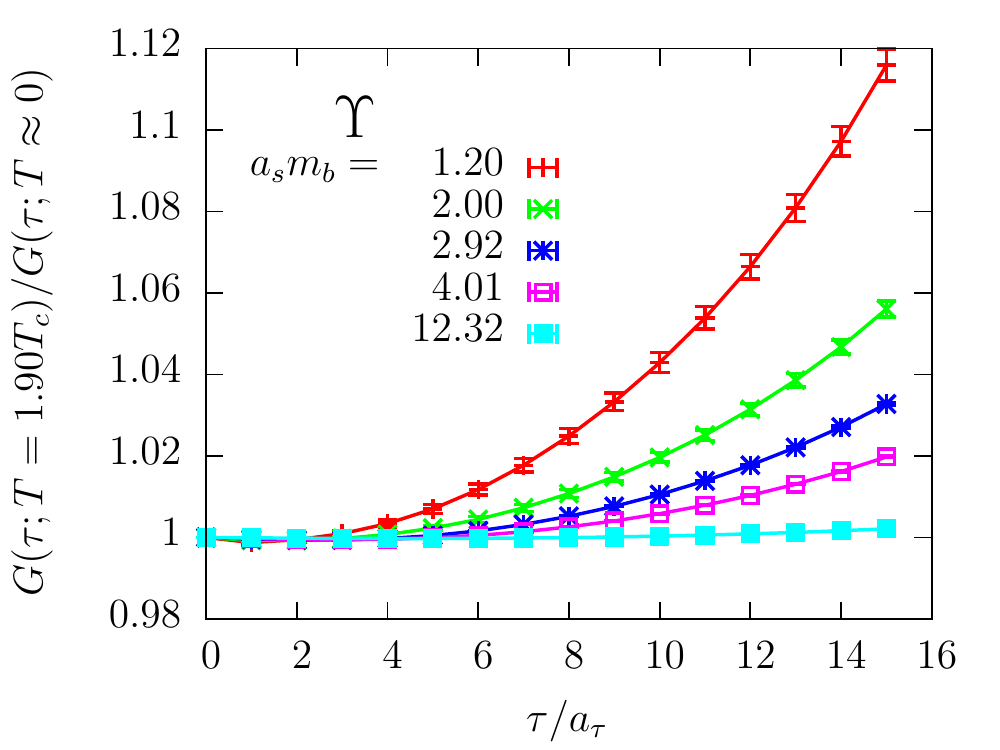}%
        \end{subfigure}
        \begin{subfigure}[t]{0.5\textwidth}
            \includegraphics[width=\textwidth]{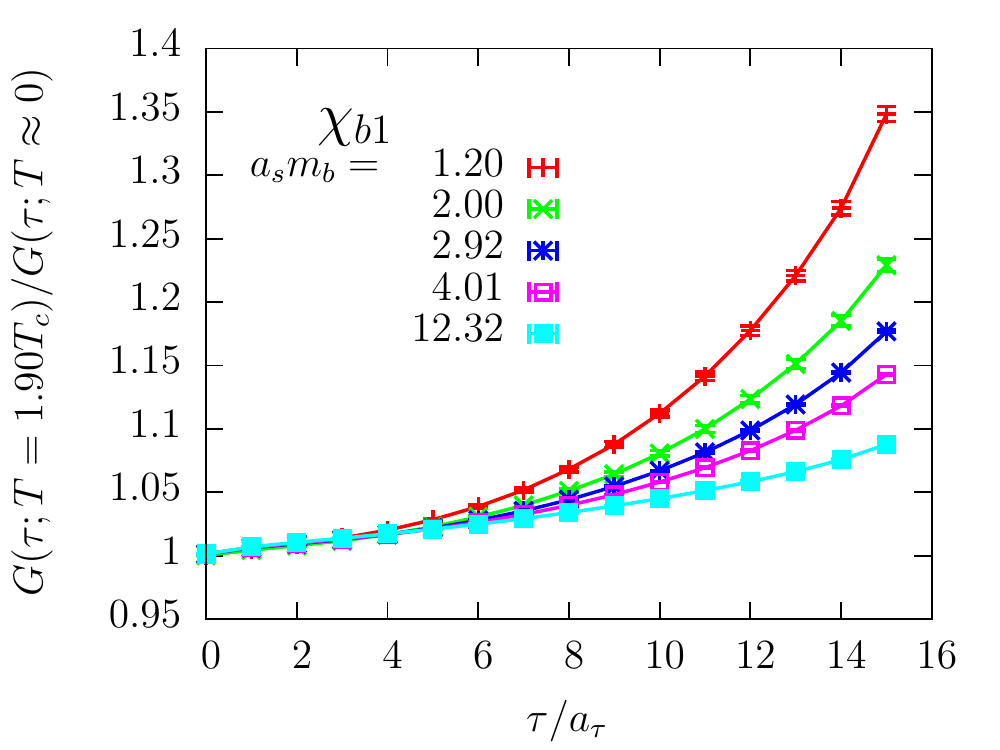}%
        \end{subfigure}
    }\vskip -1em
    \caption{Dependence on the heavy quark mass of the modification in the correlators at the highest accessible temperature, $T/T_c=1.90$, in the $\Upsilon$ (left) and $\chi_{b1}$ (right) channels.}
    \label{fig:mbdep}
\end{figure}
The thermal modification of the spectrum is expected to depend on the heavy quark mass.
In simple potential models the binding radius is related to the typical inverse momentum transfer, $r_H\sim(m_bv)^{-1}$, and effective colour Debye screening occurs when the screening length, $r_D$, is comparable with or smaller than the binding radius, $r_D\lesssim r_H$. 
At a given temperature, we would therefore expect such a mechanism to become less effective for heavier, more tightly bound states.
Figure~\ref{fig:mbdep} shows the modification of correlators for various lattice heavy quark masses and illustrates the mass dependence in the $\Upsilon$ (left) and $\chi_{b1}$ (right) channels.
Correlators in the $\chi_{b1}$ channel exhibit greater thermal modification than in the $\Upsilon$ channel at each of the lattice heavy quark masses investigated.
At smaller values of the heavy quark mass, approaching the charm quark mass, a large enhancement is seen even in the $\Upsilon$ channel correlation function, while for large values of the heavy quark mass some enhancements are still seen in the $\chi_{b1}$ channel.
The mass dependence has also been investigated in the $N_f=2$ case~\cite{Kim:2012by}.
\subsection{Spectral functions}
\label{sec:specfns}
\begin{figure}[t]
    \centering{
            \includegraphics[scale=0.9]{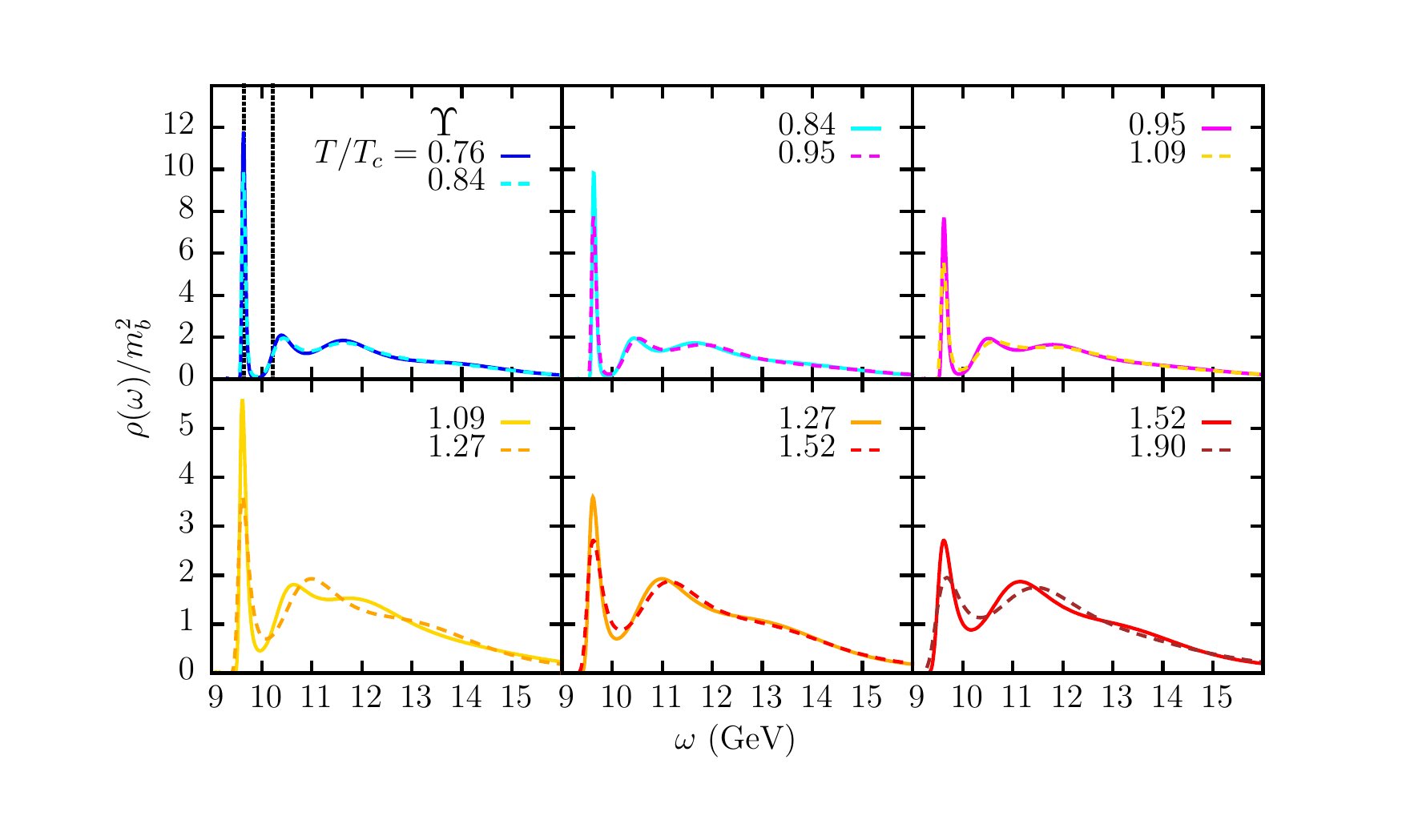}
            \vskip -2em
            \caption{Temperature dependence of the reconstructed spectral function in the $\Upsilon$ channel.
            The dashed black lines in the first panel indicate the ground state and first excited state energies determined from multi-exponential fits at zero temperature.
            Note the different ordinate scale between the upper and lower panels.}
            \label{fig:upsilon}
        }
\end{figure}

\begin{figure}[ht]
    \centering
            \includegraphics[scale=0.9]{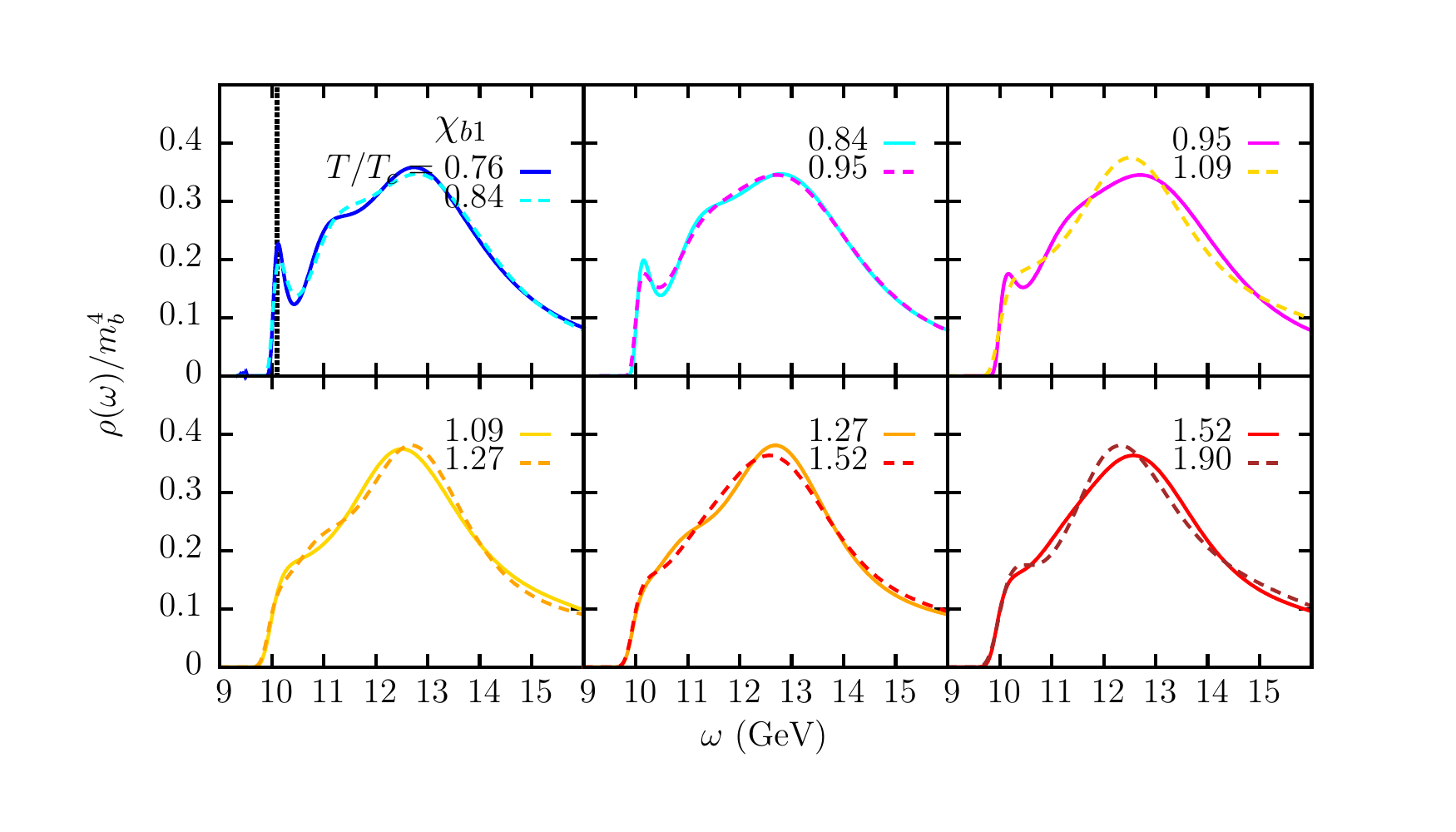}
            \vskip -2em
            \caption{Temperature dependence of the reconstructed spectral function in the $\chi_{b1}$ channel with the zero temperature ground state energy shown in the first panel with a dashed black line.}
            \label{fig:chi1}
    
\end{figure}

Figures~\ref{fig:upsilon} and \ref{fig:chi1} depict the spectral functions in the $\Upsilon$ and $\chi_{b1}$ channels respectively at temperatures from $0.76T_c$ up to $1.90T_c$.
For clarity each panel displays just two neighbouring temperatures.
In the $\Upsilon$ channel the ground state peak is clearly visible and coincides with the energy extracted from the exponential fit to the correlation function at zero temperature, see figure~\ref{fig:rho_16x128} (left).
The ground state peak persists at all accessible temperatures demonstrating the survival of the ground state to at least $T=1.90T_c$.
We observe a broadening in the peak and a decrease in its height above $T_c$.
Below $T_c$ the second peak may be identified with the first excited state.
Its interpretation above $T_c$ is less clear which may be due to melting as well as the possible presence of lattice artefacts in the high frequency part of the spectral function, which are discussed further in appendix~\ref{sec:app}.
In the $\chi_{b1}$ channel the ground state peak can be discerned at temperatures below the $T_c$ and agrees with the energy from the exponential fit at zero temperature.
This peak is observed to disappear immediately in the deconfined phase which we suggest indicates the dissociation of this state almost as soon as the deconfined phase is reached.
We note that the ground state peak in the P~wave channels is harder to distinguish than in the S~wave channels, even below $T_c$.
\section{Systematic tests of MEM}
\label{sec:systests}
\begin{figure}[t]
    \centerline{
        \includegraphics[scale=0.9]{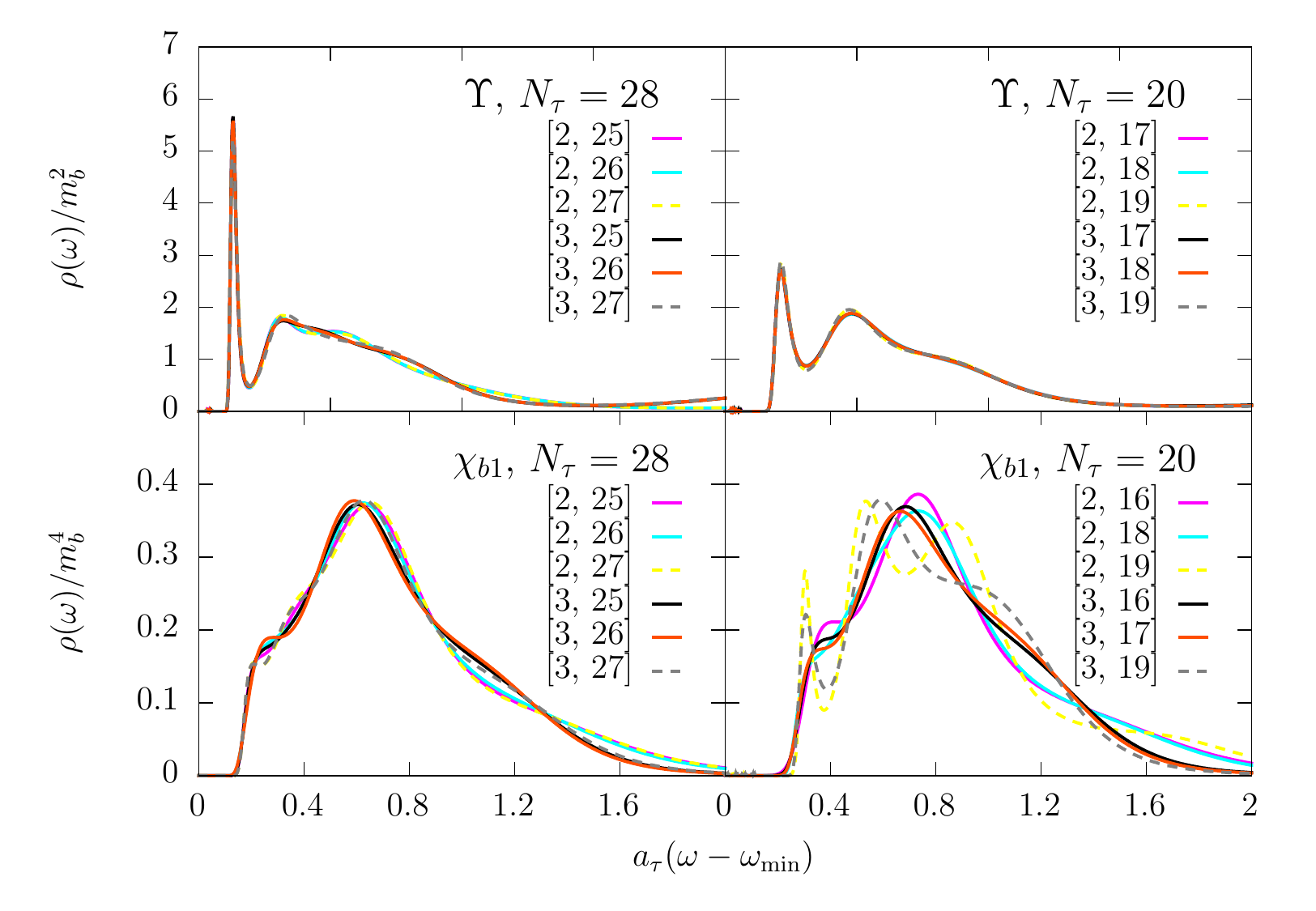}
    }
    \caption{Stability of the reconstructed spectral function with respect to selection of temporal correlator data $[\tau_1/a_\tau,\,\tau_2/a_\tau]$ shown in the key for the $\Upsilon$ (top) and $\chi_{b1}$ (bottom) channels, for two temperatures corresponding to $N_\tau=28$ (left) and $N_\tau=20$ (right).
The results when the largest temporal separation ($\tau/a_\tau=N_\tau-1$) is included are shown with dashed lines.}
    \label{fig:t1t2stab}
\end{figure}

\begin{figure}[ht]
    \centering
    \includegraphics[scale=0.9]{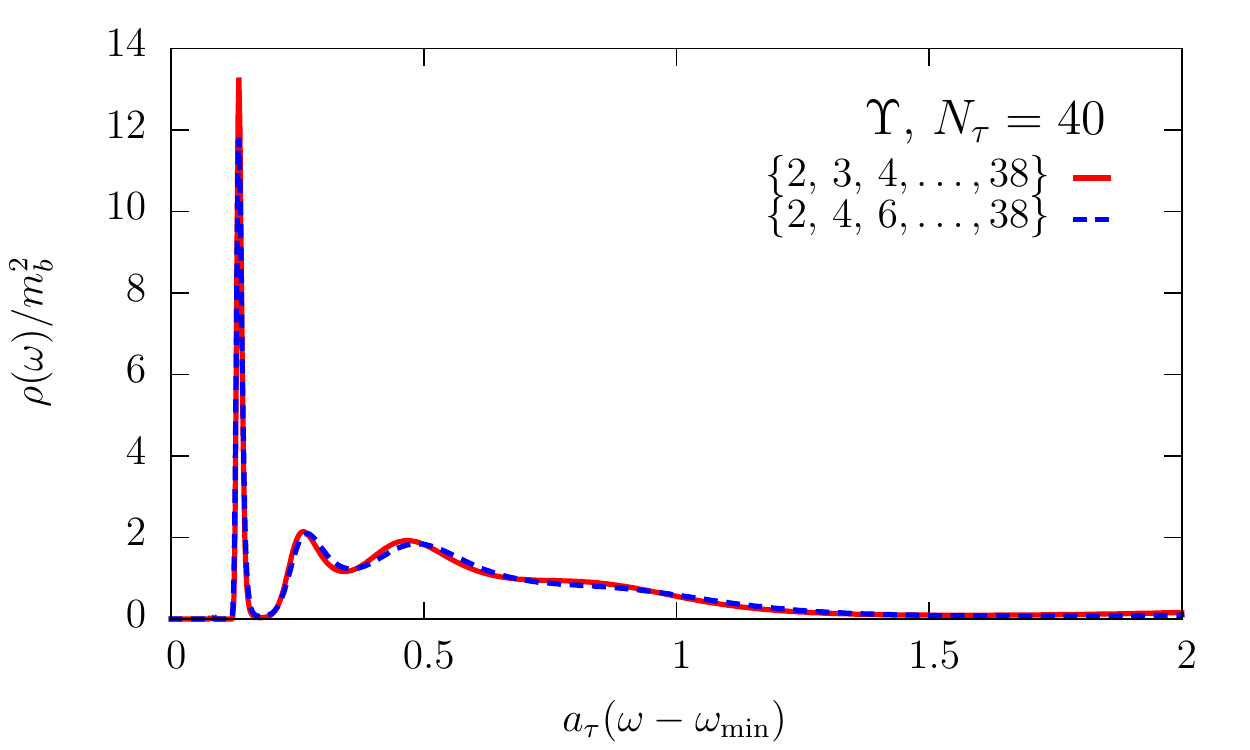}
    \caption{Reconstructed spectral function in the $\Upsilon$ channel using a subset of the correlator data compared with using the full correlator data.}
    \label{fig:half_tslice}
\end{figure}

A close examination of the reconstruction of the spectral functions is essential to have confidence in the interpretation of temperature effects.
Here we address some pertinent issues due to the selection of the temporal range of the correlator and the frequency domain of the spectral function used in the MEM.
Other effects such as the dependence on the default model and the statistical uncertainty have been investigated for similar data from the $N_f=2$ ensembles~\cite{Aarts:2011sm,Aarts:2013kaa} where they were noted to be mild.
Typical systematic effects in lattice studies such as the unphysical pion mass and finite-volume effects are not discussed although the latter are expected to be small for such heavy quarkonium states.
The stability of the spectral function with the variation of the temporal range of the correlation functions used in the reconstruction is shown in figure~\ref{fig:t1t2stab}.
The spectral functions are stable as long as data at temporal separations close to $\tau/a_\tau=0$ or $N_\tau$ are excluded on account of lattice artefacts.
Effects due to the inclusion of temporal separations near $N_\tau$ have also been discussed in ref.~\cite{Aarts:2013kaa}.
Although there are no temporal boundary conditions for the heavy quark fields, we recall that the gauge fields are periodic.
Since the spatial lattice spacing is coarser than the temporal one, $a_s=3.5a_\tau$, effects at separations close to $N_\tau$ may be expected at this scale.
The effect is stronger in the P wave channels~\cite{Aarts:2013kaa}.
In the $\Upsilon$ channel (figure~\ref{fig:t1t2stab}, top panels) the spectral function is stable when varying the temporal window as long as the correlator datum closest to $\tau/a_\tau=0$ is omitted. 
In the $\chi_{b1}$ channel (figure~\ref{fig:t1t2stab}, bottom panels) the spectral function is stable as long as the largest temporal separation $\tau/a_\tau=N_\tau-1$ is also excluded.
Therefore the reconstructed spectral function converges in all cases when the range of correlator data, $[\tau_1/a_\tau,\,\tau_2/a_\tau]$, is chosen such that $\tau_1/a_\tau\gtrapprox1$ and $\tau_2/a_\tau\lessapprox N_\tau-1$.
We have also investigated using a subset of the available correlator data to ensure any inferences about the temperature dependence are due to physical effects and not artefacts of the analysis using fewer data.
In figure~\ref{fig:half_tslice}, the reconstructed spectral function in the $\Upsilon$ channel at $T=0.76T_c$ using only half the available correlator data is compared with that from using the entire correlation function.
Only a small variation in the ground state peak height is observed.
The frequency domain chosen for each reconstruction of spectral function is given in table~\ref{tab:omegainterval}.
This interval must be chosen judiciously and may extend to negative frequencies as the energies can be shifted by an a priori unknown constant in the effective theory.
Furthermore, this range must be sufficiently large to exclude unphysical spectral weight which has been observed to appear at the edges of the interval.
\begin{table}[ht]
    \centering
    \begin{tabular}{*{3}{c}}
        \toprule
        & $\Upsilon$                                                      & $\chi_{b1}$                                                      \\
                     $N_\tau$ & $a_\tau\omega_{\textrm{min}},\,a_\tau\omega_{\textrm{max}}$ & $a_\tau\omega_{\textrm{min}},\,a_\tau\omega_{\textrm{max}}$ \\
                     \midrule
                     128       & \phantom{-}0.12, 2.12                                       & \phantom{-}0.18, 2.18                                       \\
                     40       & \phantom{-}0.08, 2.08                                       & \phantom{-}0.16, 2.16                                       \\
                     36       & \phantom{-}0.08, 2.08                                       & \phantom{-}0.16, 2.16                                       \\
                     32       & \phantom{-}0.08, 2.08                                       & \phantom{-}0.16, 2.16                                       \\
                     28       & \phantom{-}0.08, 2.08                                       & \phantom{-}0.10, 2.10                                       \\
                     24       & \phantom{-}0.08, 2.08                                       & \phantom{-}0.08, 2.08                                       \\
                     20       & \phantom{-}0.00, 2.00                                       & \phantom{-}0.00, 2.00                                       \\
                     16       & -0.04, 1.96                                                 & -0.04, 1.96                                                 \\
        \bottomrule
    \end{tabular}
    \caption{Frequency ranges used in reconstruction of the spectral functions.
                The frequency interval is discretized into $N_\omega=1000$ points for each $N_\tau$.}
    \label{tab:omegainterval}
\end{table}
\section{Conclusions}
\label{sec:concs}
Calculating heavy quarkonium correlation functions using lattice QCD provides valuable input towards understanding the modification of the spectrum in the hadronic and plasma phases of QCD.
We observe that with increasing temperature the correlation functions are enhanced relative to low temperature.
In NRQCD this temperature dependence arises solely from the temperature dependence of the associated spectral function since the integral kernel relating the spectral function and the correlator is temperature independent.
These enhancements are greater in the P~wave than the S~wave channels for each temperature below and above $T_c$.
There is significant mass dependence in these modifications, with lighter states showing increased temperature dependence.
Further interpretation is aided by calculating the spectral functions using MEM.
At zero temperature the reconstructed spectral functions have localised peaks coincident with bound state energies extracted directly from the correlation functions.
The analysis of the spectral functions at finite temperature suggests the survival of the ground state $\Upsilon$ up to at least $1.90T_c$ with some modification above $T_c$.
The $\Upsilon(2S)$ state appears to dissolve close to $T_c$, although the proximity of lattice artefacts complicates the interpretation of this structure.
The ground state $\chi_{b1}$ peak is suppressed immediately above the crossover temperature indicating significant alterations, compatible with the dissociation of this state in the QGP.
These results are qualitatively consistent with the conclusions of the earlier $N_f=2$ studies and a systematic comparison between the temperature dependence of the peak positions and widths is underway.
In the future we also plan to investigate the momentum dependence of the bottomonium spectral functions and examine new Bayesian approaches to the reconstruction of the spectral functions~\cite{Kim:2013seh,Burnier:2013nla}.
\acknowledgments
We thank D.~K. Sinclair for code used in the calculation of the heavy quark propagators and A. Rothkopf for stimulating discussions about the Bayesian reconstruction of the spectral functions.
We are grateful to the Hadron Spectrum Collaboration for the use of their zero temperature ensemble.
This work is undertaken as part of the UKQCD collaboration and the STFC funded DiRAC Facility.
We acknowledge the PRACE Grants 2011040469 and Pra05 1129, the Initial Training Network STRONGnet, Science Foundation Ireland grants 11-RFP.1-PHY3193 and 11-RFP.1-PHY-3201, the Trinity Centre for High-Performance Computing, the Irish Centre for High-End Computing, STFC, the Wolfson Foundation, the Royal Society and the Leverhulme Trust for support.
SK is supported by the National Research Foundation of Korea, grant No. 2010-002219.
    
\begin{appendices}
    \section{Free lattice spectral functions}
    \label{sec:app}
\begin{figure}[ht]
    \centering
    \includegraphics[scale=1.0]{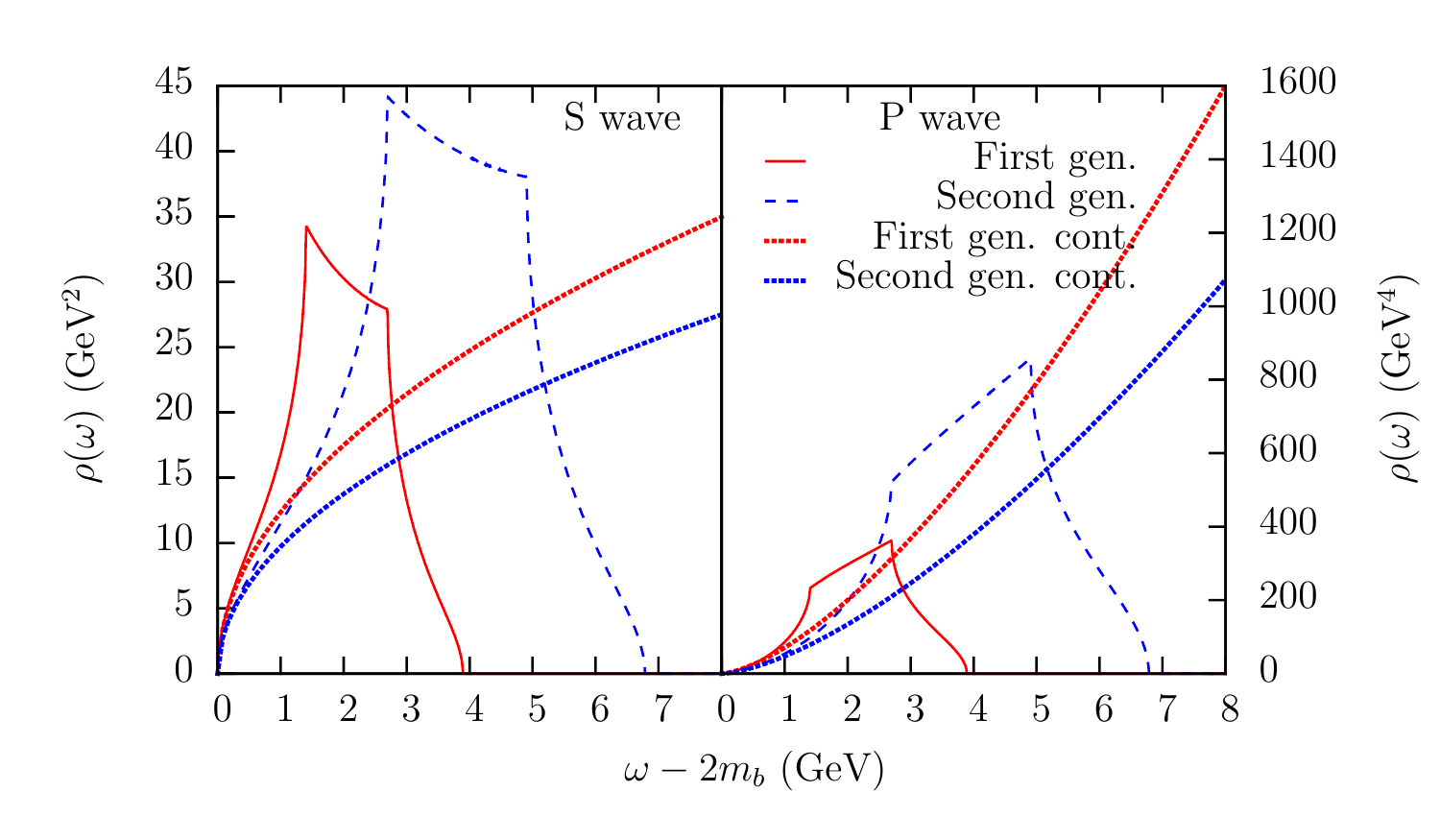}
    \caption{Free lattice spectral functions in a large volume for the second generation parameters (in blue) and the first generation parameters (in red) for both the S~wave (left panel) and P~wave (right panel).}
    \label{fig:rhofree}
\end{figure}

    The free lattice spectral functions are calculated by summing over the first Brillouin zone~\cite{Aarts:2011sm} according to
\begin{align}
    a_s^2\rho_{\mathrm S}(\omega) &= \frac{4\pi N_c}{\xi N_s^3}\sum_{\bm n\in\textrm{1BZ}}\delta(a_\tau\omega-2a_\tau E(\bm n)), \\
    a_s^4\rho_{\mathrm P}(\omega) &= \frac{4\pi N_c}{\xi N_s^3}\sum_{\bm n\in\textrm{1BZ}}\hat p^2 \delta(a_\tau\omega-2a_\tau E(\bm n)).
    \label{}
\end{align}
The lattice dispersion relation corresponding to the improved NRQCD action is given by
\begin{align}
    a_\tau E(\bm n) = 
    -2\log\left(1 - \frac{1}{2}\frac{\hat p^2}{2\xi a_sm_b}\right)
    -\log\left(1 - \frac{\hat p^4}{24 a_sm_b \xi}
    +\left(1 + \frac{a_sm_b}{2\xi}\right)\frac{(\hat p^2)^2}{8\xi (a_sm_b)^3}\right)
    \label{eq:disprel}
\end{align}
with the lattice momenta defined by
\begin{align}
    \hat p^2 = 4\sum_{i=3}^3\sin^2\left(\frac{\pi n_i}{N_s}\right),\quad
    \hat p^4 = 16\sum_{i=3}^3\sin^4\left(\frac{\pi n_i}{N_s}\right),\quad
    n_i = -\frac{N_s}{2}+1,\ldots,\frac{N_s}{2}.
   \label{eq:latdisp}
\end{align}
A comparison between the free lattice spectral functions for the first and second generation parameters (see table~\ref{tab:latparam}) is shown in figure~\ref{fig:rhofree} for the S~wave (left) and P~wave (right) channels.
The effect of the finer spatial lattice spacing is apparent as the cusp artefacts --- which correspond to momenta reaching the corners of the Brillouin zone --- are pushed to higher frequencies, and the support of the free lattice spectral function is correspondingly enlarged.
Note that at high temperatures the support of the reconstructed spectral functions (figures~\ref{fig:upsilon} and~\ref{fig:chi1}) is comparable with that of the free lattice spectral function.
{The difference between the dotted lines representing the free spectral functions in the continuum is due to the slightly different choice of heavy quark mass between the two ensembles.}
\end{appendices}
\bibliographystyle{JHEP-2}
\bibliography{2ndGenNRQCD}
\end{document}